\documentstyle[12pt]{article}
    \def\setfonts{%
  \font\frbig=eufm10 scaled \magstep1 
  \font\frscr=eufm10
  \font\frscrscr=eufm8
  \newfam\frfam
  \textfont\frfam=\frbig
  \scriptfont\frfam=\frscr
  \scriptscriptfont\frfam=\frscrscr
  \def\fr{\fam\frfam}

  \font\openbig=msbm10 scaled\magstephalf
  \font\openscr=msbm8 
  \font\openscrscr=msbm8
  \newfam\openfam
  \textfont\openfam=\openbig
  \scriptfont\openfam=\openscr
  \scriptscriptfont\openfam=\openscrscr
  \def\open{\fam\openfam}
  }

\setfonts

\def\oC{{\open C}}

\def\oZ{{\open Z}}
\def\oR{{\open R}}
\def\oV{{\open V}}

 \def\oN{{\open N}}

\begin{document}
\baselineskip=20pt

\begin{center}
{\Large\bf
SPECTRAL PROPERTIES OF WICK POWER SERIES OF A FREE FIELD WITH AN INDEFINITE
METRIC}

\bigskip
{\large
A.~G.~Smirnov$^{_1}$, M.~A.~Soloviev}\footnote{
Lebedev Physics Institute, RAS, Moscow, Russia.
}
\end{center}

\vspace{0.3cm}

\begin{center}
{\large
Abstract}
\end{center}

\bigskip
The properties of infinite series in the Wick powers of a free field whose
two-point correlation function has a singular infrared behavior
and does not satisfy the positivity condition are investigated. If these
series are defined on an appropriate functional domain, then the fields to
which they converge satisfy all conditions of the pseudo-Wightman
formalism. For series convergent only on analytic test functions in the
momentum representation, the spectral condition is formulated using the
previously introduced notion of a carrier cone of an analytic functional. A
suitable generalization of the Paley-Wiener-Schwartz theorem is used to
prove that this condition is satisfied.

 \section{ Introduction}

In \cite{SmirnovSoloviev} a criterion was established that characterizes
test functions such that averaging with these functions ensures the
convergence of an infinite series in the Wick powers of a free field in a
theory with an indefinite metric. We show here that if the functional
domain for Wick series is so defined, then the fields to which they
converge satisfy all conditions of the pseudo-Wightman formalism
\cite{BLOT}. We also indicate a refinement of this formalism whose
necessity follows from the analysis of the problem. The major difficulty
lies in the correct statement of the spectral condition in the case when
the representation of the translation group is pseudounitary and the series
converges only on analytic test functions in the momentum representation,
for which the usual formulation in terms of the spectral measure support or
the distribution support becomes impossible. It is important that this case
is rather typical than exceptional. For example, it includes the situation
with the normal exponential of the dipole ghost field
\cite{MoschellaStrocchi, Moschella}, which enters into the exact solutions
of some gauge models. The problem of finding the corresponding
generalization of the spectral condition was raised in
\cite{MoschellaStrocchi}; it was suggested in \cite{Soloviev1} that it
should be solved using the notion of a carrier cone (quasisupport) of an
analytic functional, which had been successfully applied in nonlocal field
theory. The results presented below show that this approach provides an
exact description of the spectral properties of Wick series and can serve
as a starting point for constructing a consistent Euclidean formulation of
field theory with indefinite metric.

In Sec. 2, the properties of Wick series are analyzed in the traditional
formalism, which applies if the functional domain of the series contains
functions of compact support in both the coordinate and momentum
representations. In Sec. 3, some mathematical tools required for analyzing
the general case are introduced. To this end, we present certain facts
concerning the Gelfand-Shilov spaces $S_a^b$ that generalize the more
frequently used spaces $S_\beta^\alpha$, which were introduced by the same
authors. The theory of the spaces $S_a^b$ was only outlined in
\cite{GelfandShilov}; here, we derive some new theorems for them which are
needed for solving the problem under study and are also of independent
interest. In Sec. 4, the Laplace transformation of analytic functionals on
$S_a^b$ is considered. In Sec. 5, the precise formulation of the
generalized spectral condition is given, and the proof is presented that
it is satisfied for the sums of Wick series. The main tool in the proof is
a Paley-Wiener-Schwartz-type theorem established in Sec. 4 for analytic
functionals with an acute carrier cone. Section 6 is devoted to concluding
remarks.

 \section{Properties of Wick series in the traditional formalism}

Let $\phi$ be a neutral scalar free field acting in a pseudo-Hilbert state
space ${\cal H}$, and let $d_k$ be the coefficients of a series in the Wick
powers $:\phi^k:$ of this field. Let $D^E$ denote the linear span of the
vacuum $\Psi_0$ and all vectors of the form
$$
\prod_{j=1}^n:\phi^{k_j}:(f_j)\Psi_0
$$
where the test functions $f_j$ run over a dense subspace $E$
in the Schwartz space $S$.
It was proved in \cite{SmirnovSoloviev} that under some natural conditions
on $d_k$ and for a suitably chosen $E$, the sequence of partial sums
$$
\varphi_N(f)=\sum_{k\le N} d_k:\phi^k:(f)\qquad (f\in E)
$$
with the domain $D^E$ has a strong graph limit $\varphi(f)$
which is an operator-valued distribution over $E$. Moreover, any Wick
series subordinate to the one in question in the sense that its
coefficients $d'_k$ satisfy the inequality
$|d_k^{\prime}|\le C\, |d_k|$ is convergent, and all fields determined by
such series have a common dense and invariant domain
$D^E(\varphi)\subset {\cal H}$. Since the operator realization is
constructed directly in the state space of the original free field $\phi$,
a part of the conditions of general quantum field theory \cite{BLOT} hold
for the field $\varphi$ in an obvious way. In particular, in view of the
definition of strong graph limit \cite{ReedSimon}, it follows from the
mutual locality of Wick monomials that the field $\varphi$ is local and,
moreover, mutually local with any field $\varphi'$ determined by a
subordinate series, i.e.,
$$
      [\varphi(f),\varphi'(f')]\,\Psi =0
$$
for any fixed test functions $f,f'\in E$ with spacelike separated supports
and for all $\Psi\in D^E(\varphi)$. Originally, there is a pseudounitary
representation of the Poincar\'e group in the space ${\cal H}$. Its
implementers $U(\xi,\Lambda)$ are defined on the cyclic domain $D_0$ of the
field $\phi$ and transform it into itself. As is shown by simple examples,
these operators are not necessarily bounded, but since they are
pseudounitary, they are closable. By the construction of the Wick
monomials, their closures are defined on the subspace $D^S$ and the more so
on $D^E$. Let us consider a sequence
$U(\xi,\Lambda)\varphi_N(f)U(\xi,\Lambda)^{-1}\Psi_0$, where $f\in E$.
According to the transformation law for Wick monomials, which follows from
that for $\phi$, this sequence coincides with
$\varphi_N(f_{(\xi,\Lambda)})\Psi_0$, where
$f_{(\xi,\Lambda)}(x)=f(\Lambda^{-1}(x-\xi))$. Passing to the limit as
$N\to\infty$ and taking into account that the operator $U(\xi,\Lambda)$
is closable and the vacuum is invariant under it, we see that this operator
can be extended to the vectors of the form
$\varphi(f)\Psi_0$ in a one-to-one way and that it transforms them into
$\varphi(f_{(\xi,\Lambda)})\Psi_0$. Similarly, we have
$$
      U(\xi,\Lambda)\varphi(f)
 \prod_{j=1}^n\varphi(f_j)\Psi_0 = \varphi(f_{(\xi,\Lambda)})
 \prod_{j=1}^n\varphi(f_{j(\xi,\Lambda)})\,\Psi_0
$$
for test functions in $E$. As a consequence, the relation
$$
U(\xi,\Lambda)\varphi(f)\, U(\xi,\Lambda)^{-1}\Psi=
\varphi(f_{(\xi,\Lambda)})\,\Psi
$$
holds for any $\Psi\in D^E(\varphi)$, i.e., the Poincar\'e covariance
condition is satisfied.

Let ${\tt d}$ be the space-time dimension. One of the natural conditions on
the space $E$ is that the tensor product $E(\oR^{\tt d})^{\otimes n}$
should be dense in $E(\oR^{n{\tt d}})$. In this case, the multilinear form
$\Psi(f_1,\ldots,f_n)=\varphi(f_1)\ldots\varphi(f_n)\Psi_0$
uniquely determines a vector-valued distribution on
$E(\oR^{n{\tt d}})$. Indeed, let $f\in E(\oR^{n{\tt d}})$, let $f_\nu\in
E(\oR^{\tt d})^{\otimes n}$, and let $f_\nu\to f$ as
$\nu\to\infty$. We have the relation
\begin{equation}
   \|\Psi(f_\nu)-\Psi(f_\mu)\|^2=
T_n\left((\overline{f_\nu-f_\mu})\otimes(f_\nu-f_\mu)\right),
\label{lab1}
\end{equation}
where the distribution $T_n$ can be expressed in terms of the two-point
Wightman function of the field $\phi$ and its Hilbert majorant (see
\cite{SmirnovSoloviev}). The space $E$ is chosen in \cite{SmirnovSoloviev}
in such a way that $T_n$ is defined and continuous on $E(\oR^{2n{\tt d}})$.
Therefore $\|\Psi(f_\nu)-\Psi(f_\mu)\|\to 0$ as $\nu,\mu\to\infty$.
Since the space ${\cal H}$ is complete, the Cauchy sequence $\Psi(f_\nu)$
converges in it to a vector, for which we use the standard notation
\begin{equation}
 \Psi(f)=\int\varphi(x_1)\ldots\varphi(x_n)\,f(x_1,\ldots,x_n)\,
 {\rm d}x_1\ldots {\rm  d}x_n\Psi_0.
\label{lab2}
\end{equation}
The mapping $E(\oR^{n{\tt d}})\to {\cal H}$ thus defined is linear and
continuous. The representation $U(\xi,\Lambda)$ satisfies the
spectral condition
\begin{equation}
{\rm supp}\,\int\langle\Phi,\,U(\xi,I)\Psi\rangle\,e^{-ip\xi}\,
{\rm d}\xi\subset \bar\oV_+ ,
\label{lab3}
\end{equation}
where $\oV_+$ is the upper light cone and the bar denotes closure.
Here we originally deal with $\Phi, \Psi\in D_0$, but by the construction
of Wick monomials, the condition (\ref{lab3}) remains valid for the vectors in $D^S$
as well, and the matrix element under the integral sign remains
polynomially bounded with respect to $\xi$ in this case.
However, we stress that this boundedness property does not necessarily hold
for {\it arbitrary} $\Phi$ and $\Psi$ belonging to the domain of the
closure of the operator $U(\xi,I)$.
If the Fourier transform ${\cal F}(E)$ of $E$ contains functions of
compact support, then the spectral properties of the functional
$\Psi(f)$ defined by (\ref{lab2}) can be derived from (\ref{lab3}) in the usual way.
Namely, let $\Phi\in D^E(\varphi)$. The expression
$\langle\Phi,\,U(\xi,I)\Psi(f)\rangle$ is the convolution of the test
function $f$ and a distribution in $E'$ (considered in the plane
$x_1=\ldots=x_n=\xi$), and therefore its growth with respect to $\xi$
is limited by the indicator function defining the space $E$. Let
$f_1\in E(\oR^{\tt d})$. Then
\begin{equation}
\int\langle\Phi,\,U(\xi,I)\Psi(f)\rangle\, f_1(\xi)\,{\rm  d}\xi=
\langle\Phi,\,\Psi(f_2)\rangle,
\label{lab4}
\end{equation}
where
$$
f_2(x_1,\ldots,x_n)=\int f(x_1-\xi,\ldots,x_n-\xi)\,f_1(\xi)\,{\rm  d}\xi.
$$
The Fourier transform takes $f_2$ to
$$
\hat f_2= \int   e^{ipx}f_2(x)\,{\rm d}x=
\hat f\, \hat f_1 (p_1+\ldots+p_n).
$$
According to (\ref{lab3}) the integral (\ref{lab4})
vanishes if ${\rm supp}\,\hat f_1\cap \bar\oV_+ =\emptyset$.
On the other hand, we have
$\Psi(f)\stackrel{\rm def}{=}\hat\Psi(\check f)$, where
$$
\check f(p)=(2\pi)^{-n{\tt d}}\int e^{-ipx}f(x)\,{\rm d}x.
$$
Consequently, the support of $\hat\Psi$
with respect to the variable $p_1+\ldots+p_n$ is contained in $\bar\oV_-$.
The same is true for the variable $p_m+\ldots+p_n$, $m\leq n$,
i.e., the functional (\ref{lab2}) in the momentum space is supported by the cone
\begin{equation}
K_{n-}=\{p\in \oR^{{n}{\tt d}}:  p_m+\ldots+p_n\in \bar\oV_-,\quad
\forall\,\, m=1,\ldots, n\}.
\label{lab5}
\end{equation}
In particular, this cone contains the support of the Fourier transform
of the $n$-point Wightman function of the field $\varphi$, which is
clear from its explicit expression via the two-point function of the
original field $\phi$. If ${\cal F}(E)$ consists of analytic functions,
then the above derivation is no longer valid, and some other means of
functional analysis are needed for describing the spectral properties.

\section{Gelfand-Shilov-Gurevich spaces}
It was shown in \cite{SmirnovSoloviev} that a suitable space $S_a^b$ can be
taken as the functional domain $E$ of the Wick series $\sum
d_k:\phi^k:(x)$. Such spaces are particularly convenient from the
standpoint of the Fourier transformation, which simply interchanges the
indices $a$ and $b$. The defining indices are sequences of positive numbers
satisfying the regularity conditions
\begin{equation}
a_{k+l}\leq C_1h_1^{k+l}a_ka_l, \quad b_{k+l}\leq
C_2h^{k+l}_2b_kb_l,
\label{lab6}
\end{equation}
where $C_{1,2}$ and $h_{1,2}$ are constants
\footnote{The conditions (\ref{lab6}) ensure that the operations required
for a convenient calculus are performable in $S^b_a$. It is noteworthy that
the weaker conditions $a_{k+1}\leq C_1h_1^k a_k, \quad
 b_{k+1}\leq C_2 h^k_2 b_k$ proved to be sufficient for deriving the
 results in \cite{SmirnovSoloviev}.}. The space $S_b^a={\cal
 F}(S^b_a)$, which in the context of the problem under study is the
 test function space in the momentum representation,
consists of smooth functions satisfying the bounds
\begin{equation}
|p^\lambda\partial{\,^\kappa} g(p)|\leq
  CA^{|\kappa|}B^{|\lambda|} a_{|\kappa|}b_{|\lambda|},
\label{lab7}
\end{equation}
where $\kappa$ and $\lambda$ are multi-indices whose norm is defined
as the sum of the components and the constants $A$, $B$ and $C$
depend on $g$. The conditions (\ref{lab7}) can be rewritten as
\begin{equation}
b(|p|/B)|\partial{\,^\kappa} g(p)|\leq
  CA^{|\kappa|}a_{|\kappa|},
\label{lab8}
\end{equation}
where $b(s)=\sup_{l\in \oN}(s^l/b_l)$, $|p|=\max_j|p_j|$, and $\oN$
denotes the set of nonnegative integers.
The function $b(s)$ is called the indicator function. If
$a_k^{1/k}={\it O}(k)$, then the elements of $S_b^a$
are analytic, and, as a rule, the indefinite metric Wick series converge
only on such test functions. This means that the matrix elements of the
fields determined by these series are analytic functionals in the momentum
representation. A special role is played by the spaces defined by the
sequence $a_k=k!$, which are usually denoted by $S^1_b$.
The elements of the dual space $S^{\prime 1}_b$
can be interpreted as hyperfunctions increasing at infinity no faster
than $C_\epsilon b(\epsilon |p|)$, where $\epsilon$ is arbitrarily small.
As is known, hyperfunctions form the broadest class of distributions
for which the notion of support is well defined. It was shown in
\cite{Soloviev2,Soloviev3} that a part of their properties are
inherited by analytic functionals of the class $S^{\prime a}_b$, where
$a_k=k^{\alpha k}$, $b_l=l^{\beta l}$ and the numbers $\alpha$ and $\beta$
satisfy the conditions $0\leq\alpha<1$ and $\beta>1$. Here, we extend
the theory developed in \cite{Soloviev2,Soloviev3} to a broader class of
functionals.

{\bf Definition 1.} Let $\alpha(s)$ and $\beta(s)$ be nonnegative
continuous functions indefinitely increasing on the half-axis
$s\geq 0$, let $\alpha(s)$ be convex and differentiable for
$s>0$, and let $\beta(s)$ be convex with respect to $\ln s$ and satisfy
the condition
\begin{equation}
2\beta(s)\leq\beta(hs)
\label{lab9}
\end{equation}
with a constant $h>1$. We define ${\cal E}_\beta^\alpha$
as the space consisting of entire analytic functions on
$\oC^n$ such that
\begin{equation}
|g(p+iq)|\leq C\exp\{\alpha(A|q|)-\beta(|p|/B)\},
\label{lab10}
\end{equation}
where the constants $A$, $B$ and $C$ depend on $g$.

We write $\alpha_1\prec\alpha$ if there are constants $C$ and $H$
such that $\alpha_1(s)\leq C+\alpha(Hs)$.

  {\bf Theorem 1.}  {\it  Let $\beta\prec \alpha$.
  Then ${\cal E}_\beta^\alpha$ coincides with the space
   $S_b^a(\oR^n)$ determined by the sequences
\begin{equation}
a_k=\sup_{r\geq 0}r^k
e^{-\alpha_*(r)}\quad {\rm и}\quad b_l=\sup_{s\geq 0}s^l e^{-\beta(s)},
\label{lab11}
\end{equation}
where $\alpha_*(r)=\sup_{s>0}(rs-\alpha(s))$.}

The function $\alpha_*$ is said to be monotonically conjugate to the convex
function $\alpha$ (see \cite{Rockafellar}). The conditions (\ref{lab6})
hold for
the sequences (\ref{lab11}) in view of (\ref{lab9}) and the inequality $2\alpha_*(s)\leq
\alpha_*(2s)$, which follows from the convexity of $\alpha_*$
 under the normalization condition $\alpha(0)=0$.

To prove Theorem 1, we need the following three lemmas.

 {\bf Lemma 1.} {\it The relation
\begin{equation}
\sup_{r\geq 0}r^k
e^{-\alpha_*(r)}=(k/e)^k\inf_{s>0}s^{-k}e^{\alpha(s)}.
\label{lab12}
\end{equation}
holds for any function $\alpha(s)$ satisfying the conditions of
Definition $1$ and for all $k\geq 0$.}

 {\bf Proof.}
The expression in the right-hand side is immediately recovered, if we
substitute $e^{-\alpha_*(r)}=\inf_{s>0}e^{\alpha(s)-rs}$
in the left-hand side of (\ref{lab12}) and change the order in which the supremum
and the infimum are calculated. Since
  $\sup_r\,\inf_s\,G(r,s)\leq\inf_s\,\sup_r\,G(r,s)$ for any function $G$,
it is only necessary to show that the opposite inequality also holds
in the case under consideration, i.e., to prove (according to the standard
terminology \cite{Rockafellar}) that the function
$G(r,s)=r^ke^{\alpha(s)-rs}$ has a saddle value.
The validity of (\ref{lab12}) with $k=0$ is obvious; we therefore suppose that
$k>0$. We recall that a differentiable convex function is continuously
differentiable. The point $s_k$ at which $\inf_{s>0} s^{-k}e^{\alpha(s)}$
is attained is determined by the equation $\alpha^{\,\prime}(s)=k/s$,
whose solution is unique because the function $\alpha^{\,\prime}$
is nonnegative, monotonic, and not identically zero. Set
   $r_k=\alpha^{\,\prime}(s_k)$. We have
\begin{equation}
\sup_{r\geq 0}\,\inf_{s>0}\,r^ke^{\alpha(s)-rs}\geq
\inf_{s>0}r_k^ke^{\alpha(s)-r_k s} =
(k/es_k)^k e^{\alpha(s_k)}
\label{lab13}
\end{equation}
because $\alpha(s)-r_k s$ is a convex function and any of its stationary
points is the point of absolute minimum. The lemma is thus proved.

{\large Lemma 2.} {\it  For any function $\beta(s)$ satisfying the
 conditions of Definition $1$, and for any $\epsilon>0$
 there is a constant $C_\epsilon$ such that}
\begin{equation}
\beta(s)+\ln s\leq C_\epsilon +\beta((1+\epsilon)s).
\label{lab14}
\end{equation}

 {\bf Proof.}
For any nondecreasing function $\beta$ which is convex with respect to
$\ln s$ and is not identically constant, there is a number
$c$ such that
\begin{equation}
\beta(s)\geq c\ln s
\label{lab15}
\end{equation}
for sufficiently large $s$. If (\ref{lab9}) holds in addition, then (\ref{lab15}) is true
for arbitrarily large $c$. The convexity with respect to
$\ln s$ implies
\begin{equation}
\beta(hs)\leq \tau\beta(h^{1/\tau}s)+(1-\tau) \beta(s)
\label{lab16}
\end{equation}
for $0\leq \tau\leq 1$. Set $\tau= N/(N+1)$ and take
$N$ such that $h^{1/N}<1+\epsilon$. Then
  $$
 \beta(hs)-\beta(s)\leq N[\beta((1+\epsilon)hs)-\beta(hs)].
 $$
Here we have $\beta(hs)-\beta(s)\geq N\ln s -C_N$
in view of the condition (\ref{lab9}) and the inequality
(\ref{lab15}) with $c=N$. After the
change of notation $hs\to s$, we obtain (\ref{lab14}).

 {\bf Lemma 3.} {\it
Let the function $\beta(s)$ satisfy the conditions of Definition $1$,
and let the sequence $b_l$ be determined by the relations $(\ref{lab11})$
and $b(s)=\sup_{l\in \oN}(s^l/b_l)$. Then
\begin{equation}
b(s)\leq e^{\beta(s)}\leq C'_\epsilon\,b((1+\epsilon)s)
\label{lab17}
\end{equation}
for any $\epsilon>0$.}

 {\bf Proof.} Only the right-hand inequality is nontrivial here.
We temporarily suppose that $l$ is an arbitrary real number,
not necessarily a positive integer, and note that the function
$b_l$ is nondecreasing for $l>\beta(e)$ because the infimum in definition
(\ref{lab11}) is attained at $s>1$ in this case. The function $\ln b_l$ is conjugate
to $\beta(e^t)$ and is finite for $l\geq 0$.
The operation of conjugation is involutory on convex closed functions
(see \cite{Rockafellar}), and therefore $\sup_{l\geq  0}(s^l/b_l)=
e^{\beta(s)}$. If $s$ is sufficiently large, then the supremum is attained
here at large $l$ because $\beta(s)$ increases faster than
$c\ln s$ with arbitrary $c$. Let $[l]$ be the integer part of $l$. Then
$$
b(s)\geq s^{[l]}/b_{[l]}\geq s^{l-1}/b_l= e^{\beta(s)-\ln(s)},
$$
and the application of (\ref{lab14}) completes the proof.

 {\bf Proof of Theorem 1.} Let $g\in {\cal
E}_\beta^\alpha$. For the natural choice of the topology on ${\cal
E}_\beta^\alpha$, the least constant $C$
for which (\ref{lab10}) is true is the norm $\|g\|_{A,B}$.
Let ${\cal D}(s)$ and $\partial_0{\cal D}(s)$ denote the polydisk
$\{\zeta\in \oC^n:\, |\zeta|=\max_j|\zeta_j|\leq s\}$ and its skeleton.
Applying the Cauchy formula, using the arbitrariness in the choice of $s$,
and taking into account the inequality
$\beta(|p|/2)\leq \beta(|\zeta+p|)+\beta(|\zeta|)$
implied by the monotonicity and nonnegativity of
$\beta$, we obtain
 $$
|p^\lambda\partial{\,^\kappa} g(p)|= |p^\lambda| \frac{\kappa !}{(2\pi)^n}
\,\left|\,\int\limits_{\partial_0{\cal
D}(s)}\frac{g(\zeta+p)}{\zeta^{\kappa+I}}\,{\rm d} \zeta\right| \leq
 \|g\|_{A,B}\,|p|^{|\lambda|}e^{-\beta(|p|/2B)}\,|\kappa| !
  \inf_{s>0}s^{-|\kappa|}e^{\alpha(As)+\beta(s/B)}
 $$
 \nopagebreak
 $$ \hspace{6.6cm}\leq
 C_\epsilon
 \|g\|_{A,B}\,(2(A+H/B+\epsilon))^{|\kappa|}(2B)^{|\lambda|}
 a_{|\kappa|}b_{|\lambda|},
 $$
where the right-hand inequality is derived
using the Stirling formula, Lemma 1,
the condition $\beta\prec\alpha$, and the inequality
$\alpha(s_1)+\alpha(s_2)\leq\alpha(2(s_1+s_2)$. Hence,
the space ${\cal E}_\beta^\alpha$ is continuously embedded in $S_b^a$.

Conversely, let $g\in S_b^a$.
The norms in $S_a^b$ will be marked with the prime.
Using the Taylor formula, the condition (\ref{lab8}), and the inequality
$|\kappa|!\leq n^{|\kappa|}\kappa!$, we obtain the estimate
$$
 |g(p+iq)|\leq
  \sum_{\kappa\in
  \oN^n}\frac{|q^\kappa|}{\kappa!}|\partial^{\,\kappa}g(p)|\leq C_\epsilon\,
 \|g\|^{\prime}_{A,B}\,\frac{1}{b(|p|/B)}\,\sup_{k\in
 \oN}(n(A+\epsilon)|q|)^k\frac{a_k}{k!}\,
$$
for the analytic continuation of $g$ into $\oC^n$. By Lemma 1 and the
inequality $1/k!\leq e^k/k^k$, we have $a_k/k!\leq e^{\alpha(s)}/s^k$ for
any~$s$. Therefore, the supremum with respect to $k$ can be majorized by
the function $e^{\alpha(n(A+\epsilon) |q|)}$. Applying Lemma 3 completes
the proof.

{\bf Remark.} It is well known that the space $S^1_b$ can be similarly
redefined. It can be regarded as being defined by the discontinuous
convex function $\alpha(s)$
vanishing for $0\leq s\leq 1$ and equal to $+\infty$ for $s>1$, and by the
function $\beta(s)=\ln b(s)$. Accordingly, its elements are analytic in
the domains of the form $|y|<1/A$.

The subclass of the spaces of type $S$ characterized by Theorem 1 will
be called the Gelfand-Shilov-Gurevich spaces because, in some respects,
they are close to the spaces of type $W$ introduced by Gurevich.
The further construction of the theory follows that in
\cite{Soloviev2,Soloviev3}, and instead of presenting complete proofs,
we comment on their main points.
The key step consists in introducing, in addition to
${\cal E}_\beta^\alpha$, the kindred spaces over cones.

    {\bf Definition 2.} Let $U$ be an open cone in ${\oR}^n$, and let
    the functions $\alpha$ and $\beta$ satisfy the conditions of Definition
1. The space ${\cal E}_\beta^\alpha(U)$ consists of entire functions on
$\oC^n$ possessing the boundedness property
\begin{equation}
|g(p+iq)|\leq
C\exp\{\alpha(A|q|)+\alpha\circ \delta_U(Ap)-\beta(|p|/B)\},
\label{lab18}
\end{equation}
where $\delta_U(p)$ is the distance from the point $p$ to the cone $U$
and the constants $A$, $B$ and $C$ depend $g$.

  {\bf Theorem 2.} {\it If the space ${\cal E}_\beta^\alpha(U)$
is endowed by the inductive-limit topology with respect to the
family of the Banach spaces
  ${\cal  E}_{\beta,B}^{\alpha,A}(U)$ of entire functions with the norms
\begin{equation}
\|g\|_{U,A,B}=\sup_{p,q}|g(p+iq)|\exp\{-\alpha(A|q|)-\alpha\circ
\delta_U(Ap)+\beta(|p|/B)\},
\label{lab19}
\end{equation}
then ${\cal E}_\beta^\alpha(U)$ becomes a {\rm DFS} space\footnote{Here DFS
is the standard notation for the class of topological vector spaces dual
  to the Frech\'et-Schwartz spaces.} and is consequently a complete,
  Hausdorff, barrelled, reflexive, separable, and Montel space.}

We recall that DFS spaces are limits of injective compact sequences of
locally convex spaces. For an arbitrary sequence contained in the unit ball
of ${\cal  E}_{\beta,B}^{\alpha,A}(U)$, the application of the Montel
theorem and the diagonalization process permit separating a subsequence
uniformly converging on compact sets in $\oC^n$ to an element of this
space. In the broader space ${\cal  E}_{\beta,B'}^{\alpha,A'}(U)$,
 $A'>A$, $B'>B$, this convergence holds with respect to the norm because
$$
\lim_{|q|\to \infty}e^{\alpha(A|q|)-\alpha(A'|q|)}=0,\quad
\lim_{|p|\to \infty}e^{-\beta(|p|/B)+\beta(|p|/B')}=0
$$
by virtue of Lemma 2. Therefore, the canonical mapping
${\cal  E}_{\beta,B}^{\alpha,A}(U) \to {\cal  E}_{\beta,B'}^{\alpha,A'}(U)$
is compact, which precisely proves Theorem 2. For the same reason,
${\cal E}_{\beta,B+}^{\alpha,A+}(U)
 =\bigcap_{A'>A,B'>B}{\cal  E}_{\beta,B'}^{\alpha,A'}(U)$ is a Frech\'et
space, which is essential for the derivation of the result below.

 {\bf Theorem 3.} {\it
For any pair $U_1,U_2\subset \oR^n$ of open cones, the relation
 $$
{\cal E}^\alpha_\beta (U_1)\mathbin{\hat{\otimes}_i} {\cal E}^\alpha_\beta (U_2)
= {\cal E}^\alpha_\beta (U_1\times U_2),
 $$
holds, where the subscript $i$ means that the tensor product is endowed
with the inductive topology and the hat denotes completion.}

The proof of this theorem is the same as that of Theorem 3 in
\cite{Soloviev3}. According to the Carleman--Ostrowski theorem on
quasianalyticity conditions, the space ${\cal F}({\cal E}^\alpha_\beta)$
contains functions of compact support if and only if
\begin{equation}
\int_1^\infty \frac{\beta(s)}{s^2}\,{\rm d}s<\infty.
\label{lab20}
\end{equation}
We assume that the condition (\ref{lab20}) holds because the formalism we develop
is intended for application to the local field theory.

 {\bf Theorem 4.} {\it Under the condition $(\ref{lab20})$, the space
 ${\cal E}_\beta^\alpha$ is dense in each of the spaces
  ${\cal E}_\beta^\alpha(U)$, where $U$ is an open cone.}

This theorem is proved in the same way as its particular case in
\cite{Soloviev3}. A much simpler proof is possible under the additional
assumption that the space ${\cal E}_\alpha^\alpha$ is nontrivial, which
suffices for the majority of applications we are interested in. Namely,
taking a function $e_0\in {\cal E}_\alpha^\alpha$
possessing the property $\int e_0(p)\,{\rm d}p=1$, we can approximate any
element $g\in {\cal E}_\beta^\alpha(U)$ with functions
$g_\nu\in {\cal E}_\alpha^\alpha$ by setting
$g_\nu=e_\nu g$, where $e_\nu$ is a sequence of Riemann sums for the
integral $\int\!e_0(p-\eta){\rm d}\eta$ or, more explicitly,
$$
e_\nu(p)=\sum_{\kappa\in \oZ^n,|\kappa|<\nu^2}e_0(p-\kappa/\nu)\nu^{-n}.
$$

A closed cone $K\subset {\oR}^n$ is called a {\it carrier cone} of a
functional $u\in {\cal E}_\beta^{\prime \alpha}$, if $u$ can be
continuously extended to every space ${\cal E}_\beta^\alpha(U)$,
where $U\supset K\setminus\{0\}$ or, in other words, if $u$ can be
continuously extended to the union
$$
{\cal E}_\beta^\alpha(K)=\bigcup_{U\supset
K\setminus\{0\}}{\cal E}_\beta^\alpha(U),
$$
endowed with the inductive--limit topology.

{\bf Theorem 5.}
{\it Any functional $u\in {\cal E}_\beta^{\prime \alpha}$
with a carrier cone $K_1\cup K_2$ can be decomposed as
 $u=u_1+u_2$, where $u_j\in {\cal E}_\beta^{\prime \alpha}(K_j),\quad
j=1,2$.}

{\bf Theorem 6.} {\it
If each of the cones $K_1$ and $K_2$ is a carrier cone of a functional
$u\in {\cal E}_\beta^{\prime\alpha}$, then $K_1\cap
K_2$ is also its carrier cone.}

As a consequence, there exists the smallest closed cone $K$ such that
$u\in {\cal E}_\beta^{\prime \alpha}(K)$. It can be called the {\it
quasisupport} of the functional $u$.

The spaces ${\cal E}_\beta^\alpha(K)$ inherit the topological properties
specified in Theorem 2. Therefore, Theorems 5 and 6 are equivalent to the
corresponding dual assertions for test functions, and among these
assertions, only the one concerning the possibility of decomposing any
function $g\in {\cal E}_\beta^\alpha(K_1\cap K_2)$ into a sum of functions
belonging to ${\cal E}_\beta^\alpha(K_1)$ and ${\cal E}_\beta^\alpha(K_2)$
needs a proof. If ${\cal E}_\alpha^\alpha$ is nontrivial, then such a
decomposition presents no problems. Namely, let
$g\in {\cal E}_{\beta,B}^{\alpha,A}(U)$, where $U$ is a conic neighborhood
of the intersection $K_1\cap K_2$, and let $W$ be its slightly smaller
neighborhood such that $\bar W\setminus\{0\}\subset U$. Since the angular
distance between the closed cones $K_j\setminus W$ is nonzero, there exist
their conic neighborhoods $W_j$ and a number $\theta>0$ such that
\begin{equation}
|p-\eta|\geq \theta |p|,\qquad |p-\eta|\geq \theta |\eta|
\label{lab21}
\end{equation}
for all $p\in W_1$ and $\eta\in W_2$. We take an arbitrary function
$e_0\in {\cal E}_{\alpha,B_0}^{\alpha, A_0}$ with the property
$\int\!e_0(p)\,{\rm d}p=1$ and set
$$
e(p)=\int_{W_2}e_0(p-\eta)\,{\rm d}\eta.
$$
Using the convexity of $\alpha$ and the inequalities (\ref{lab21}),
we can verify
that $eg\in {\cal E}_\beta^\alpha(K_1)$ given that $2B_0<\theta/A$ because
in this case, the function $e$ decreases in $W_1$ no slower than
$e^{-\alpha(A|p|)}$, which suppresses the growth of $g$ in this cone.
Similarly, we have $(1-e)g\in {\cal E}_\beta^\alpha(K_2)$
for $2B_0<\theta'/A$, where $\theta'$ is the angular distance from
$K_2\setminus W$ to the complement of $W_2$. If ${\cal E}_\alpha^\alpha$
is trivial, then the desired decomposition can be derived using the
standard method for solving Cousin problems using the H\"ormander
$L^2$-estimates (see \cite{Soloviev2}).

\section{Fourier--Laplace transformation in ${\cal E}^{\prime\alpha}_\beta$}

We proceed to extend the theory of the Fourier--Laplace transformation to
functionals of the class ${\cal E}_{\beta}^{\prime\alpha}$. The
nonquasianalyticity condition (\ref{lab20}) implies the inequality
$\beta(s)<C_\epsilon +\epsilon s$, where $\epsilon$ is arbitrarily small.
Therefore, the function
\begin{equation}
\beta^*(t)=\inf_{s>0}(st-\beta(s)).
\label{lab22}
\end{equation}
is defined on the half-axis $t>0$. We note that it is nonpositive, concave,
and monotonic and that it tends to $-\infty$ as $t\to 0$.

{\bf Теорема 7.} {\it
Let $K$ be an acute cone and $V$ be the interior of its dual cone
$K^*=\{y:py\geq 0, \, \forall p\in K\}$. If the condition $(\ref{lab20})$ holds,
then the Laplace transform ${\bf  v}(z)=(u, e^{i(\cdot,z)})$
of any functional $u\in  {\cal E}_\beta^{\prime\alpha}(K)$
is holomorphic in the domain $T^V=\{z=x+iy:\,y\in V\}$
and satisfies the inequality
\begin{equation}
|{\bf v}(z)|\leq
C_\epsilon(V')\exp\{\alpha_*(\epsilon|z|)-\beta^*(|y|/\epsilon)\}\quad
(y\in V')
\label{lab23}
\end{equation}
for an arbitrary $\epsilon>0$ and for every compact subcone $V'$ of $V$
$($i.e., such that $\bar V'\setminus\{0\}\subset V$$)$. As $y\to 0$ inside
a fixed cone $V'$, the function ${\bf v}(x+iy)$ tends to the Fourier
transform of the functional $u$ in the topology of the space ${\cal F}
({\cal E}_\beta^{\prime\alpha})$.}

 {\bf Proof.} We have
$$
|{\bf v}(z)|\leq \|u\|_{U,A,B}\|e^{i(p+iq)z}\|_{U,A,B}
$$
for any $A,B>0$ and every $U\supset K\setminus\{0\}$, where
\begin{equation}
\|e^{i(p+iq)z}\|_{U,A,B}=\sup_p\exp\{-py-\alpha\circ
\delta_U(Ap)+\beta(|p|/B)\}\, \sup_q\exp\{-qx -\alpha(A|q|)\}.
\label{lab24}
\end{equation}
In what follows, it is convenient to regard $|\cdot|$ as the Euclidean norm
in $\oR^n$. The supremum with respect to $q$ in (\ref{lab24})
with $A>1/\epsilon$
is majorized by the function $e^{\alpha_*(\epsilon|x|)}$. We take a cone
$U$ that is compact in another cone $U'$ that is compact in
${\rm int}\,V^{\prime *}$. If $p\not\in U'$, then $\delta_U(p)>\theta|p|$,
where $\theta>0$, and consequently $\alpha\circ
\delta_U(Ap)>\alpha(A\theta|p|)$. Moreover, we have $\beta(|p|/B)<C+|p|/B$,
and the supremum with respect to $p\not\in U'$ for $2A\theta>1/\epsilon$
therefore does not exceed $C_\epsilon e^{\alpha_*(\epsilon|y|)}$. If
$p\in U'$, then the inclusion $\bar U'\setminus\{0\}\subset {\rm
int}\,V'^*$ implies that $-py\leq-\theta'|p||y|$ for all $y\in V'$.
Hence, the supremum with respect to $p\in U'$ for $B\theta'>1/\epsilon$
does not exceed $-\beta^*(|y|/\epsilon)$. The estimate (\ref{lab23}) is thus proved.
It is similarly established that the difference quotients corresponding to
the partial derivatives $\partial e^{i(p+iq)z}/\partial z_j$ converge to
them for any $z\in T^V$ in the topology of ${\cal E}^{\alpha}_\beta(U)$,
and ${\bf v}(z)$ is consequently holomorphic for $z\in T^V$. By Theorem 1,
a function $f\in {\cal F}({\cal E}_\beta^{\alpha})= S_a^b$ satisfies the
inequality $|f(x)|\leq C/a(|x|/A)$, where $a(r)=\sup_k(r^k/a_k)\geq
c_\varepsilon e^{\alpha_*(r/(1+\varepsilon))}$ according to Lemma 3 applied
to $\alpha_*$. Therefore, for every $y\in V'$, the function $f$ is
integrable with the function $(u, e^{i(\cdot,z)})$ not only for the element
$u$ under consideration but also for any $u\in
 {\cal E}_\beta^{\prime\alpha}(U)$. Consequently, the sequence of Riemann
sums for the integral
$$
\int e^{iz(p+iq)}f(x)\,{\rm d}x
$$
is weakly Cauchy in the space ${\cal E}_\beta^\alpha(U)$, and because this
is a Montel space, it is convergent in ${\cal E}_\beta^\alpha(U)$. Hence,
$$
\int{\bf v}(x+iy)f(x)\,{\rm d}x= (u, e^{-(\cdot,y)}g),
$$
where
$$
g(\zeta)=\int e^{i\zeta x}f(x)\,{\rm d}x \quad (\zeta=p+iq).
$$

It remains to prove that $e^{-\zeta y}g(\zeta)\to g(\zeta)$ in the topology
of ${\cal E}_\beta^\alpha(U)$ as $y\to 0$ inside $V'$. By the first mean
value theorem, we have $|e^{-\zeta y}-1|\leq |\zeta||y|\max_{0\leq t\leq
 1}e^{-pyt}$. If $g\in {\cal E}_{\beta, B_0}^{\alpha, A_0}$, then we obtain
 $$
 \|(e^{-\zeta y}-1)g\|_{U,A,B}\leq C|y|\max_{0\leq t\leq 1}\sup_p\exp\{-pyt-
 \alpha\circ \delta_U(Ap)\}
 $$
for $A>A_0$ and $B>B_0$. Here, the expression
$-pyt-\alpha\circ \delta_U(Ap)$ in the exponential function is negative for
$p\in U'$ and $y\in V'$ and does not exceed $\alpha_*(|y|/A\theta)$ for
$p\not\in U'$. The theorem is proved.

Let $V$ be an open connected cone in $\oR^n$, and let
${\cal A}_{\alpha_*}^{\beta^*}(V)$ denote the space of holomorphic
functions in $T^V$ and satisfying the condition (\ref{lab23}). For the natural
choice of topology, it is a topological algebra with respect to
multiplication because the condition (\ref{lab9}) implies that
$2\beta^*(t)\geq \beta^*(2t/h)$. The most important for us and at the same
time the most difficult problem is to generalize the
Paley--Wiener--Schwartz theorem to analytic functionals of the class
${\cal E}^{\prime \alpha}_\beta$. We do this under an additional condition
on the growth of the elements of ${\cal A}_{\alpha_*}^{\beta^*}(V)$ at
infinity. This condition is very weak, and it holds for the applications we
are interested in. We say that two functions $\beta$ and $\beta_1$ are
equivalent if $\beta_1\prec\beta$ and $\beta\prec\beta_1$. In this case,
they specify the same space.

{\bf Theorem 8.}
{\it Let the functions $\alpha$ and $\beta$ possess the properties
specified in Definition $1$, let the conjugate function $\alpha_*$ of
$\alpha$ increase no faster than a first-order finite-type exponential
function, and let $\beta$ satisfy the condition $(\ref{lab20})$ and be equivalent to
a continuous concave function. Then the Laplace transformation establishes
an isomorphism between the space ${\cal E}^{\prime\alpha}_\beta(V^*)$
and the algebra ${\cal A}_{\alpha_*}^{\beta^*}(V)$ for any open connected
cone $V$.}

{\bf Proof.}
The cone $V$ lies inside $V^{**}$. The Laplace transform of the space
${\cal E}^{\prime\alpha}_\beta(V^*)$ is therefore contained in
${\cal A}_{\alpha_*}^{\beta^*}(V)$ by Theorem 7, and the problem consists
in proving the inverse inclusion. As before, let the sequence $a_k$ be
defined by the first formula in (\ref{lab11}). Any function ${\bf v}\in {\cal
A}_{\alpha_*}^{\beta^*}(V)$ has a boundary value in the sense of
hyperfunction theory. Under the indicated condition on the growth of
$\alpha_*$, it is defined on the space $S^1_a$, which is nontrivial under
this condition\footnote{If the function $\alpha_*$ increases so fast that
for any $N$ there exists an $r_N$ such that $\alpha_*(r)>e^{Nr}$ for
$r>r_N$, then $S^1_a$ is trivial. This follows from Theorem 4 in Chap. 2
in \cite{Mandelbrojt}.} because it contains the entire function
$\prod_{j=1}^ne^{-4\cosh z_j}$ majorizable by $e^{-ne^{|x|/n}}$ for
$|y|<\pi/3$. The elements of $S^1_a$ are analytic in the domains of the
form $\{z:\,|y|<1/B\}$ and satisfy the bound $|f(z)|\leq
\|f\|_{A,B}e^{-\alpha_*(|x|/A)}$ in them. Therefore, the formula
$$
(v,f)=\int {\bf v}(x+iy)f(x+iy)\,{\rm d}x\qquad (y\in V,\,\,|y|<1/B),
$$
where the integral is independent of $y$ by virtue of the
Cauchy--Poincar\'e theorem, defines a continuous linear functional on
$S_a^1$. We let $u$ denote the (inverse) Fourier transform of $v$ and
consider its convolution with a test function $g\in S_1^a$. We have the
identity
\begin{equation}
(u*g)(p)=(u,g(p-\cdot))=\int {\bf v}(x+iy)e^{-ip(x+iy)}f(x+iy)\,{\rm d}x,
\label{lab25}
\end{equation}
where
$$
f(z)=(2\pi)^{-n}\int g(-p)e^{-ipz}\,{\rm d}p.
$$
The Fourier transformation establishes a one-to-one and bicontinuous
mapping of $S^a_1$ onto $S_a^1$, and there exist $A'> A$ and $B'> B$ such
that $\|f\|_{S^{1,B'}_{a,A'}}\le C\,\|g\|_{S^{a,A}_{1,B}}$. Therefore,
(\ref{lab23})
and (\ref{lab25}) imply the inequality
\begin{equation}
|(u*g)(p)|\leq C_{\epsilon, A}\,\|g\|_{A,B}
\inf_{0<t<1/B'}\,e^{|p|t-\beta^*(t/\epsilon)}.
\label{lab26}
\end{equation}
Here the infimum is attained at sufficiently large $s=|p|$ on the interval
$(0,\delta)$, where $\delta$ is arbitrarily small. Indeed, suppose the
contrary, i.e., that for some $\delta>0$ and for any $N$ there exist
$s_N\geq N$ and $t_N\in [\delta,1/B')$ such that
$$
s_Nt_N-\beta^*(t_N)<\inf_{0<t<\delta}(s_Nt-\beta^*(t)).
$$
The right-hand side does not exceed $s_N\delta/2-\beta^*(\delta/2)$, and
consequently we have $N(t_N-\delta/2)<\beta^*(t_N)-\beta^*(\delta/2)$. In
view of the monotonicity of $\beta^*$, this implies the absurd inequality
$N\delta/2<\beta^*(1/B')-\beta^*(\delta/2)$. Therefore, (\ref{lab26}) remains valid
if the infimum is taken throughout the half-axis and the constant
$C_{\epsilon, A}$ is increased if necessary. The equivalence of $\beta$ to
a concave function implies its equivalence to $\beta^{**}$, which can be
verified by extending this concave function throughout the axis with the
value $-\infty$ for $s<0$ and taking into account that the operation of
conjugation is involutory for closed concave functions \cite{Rockafellar}.
Ultimately, we obtain
\begin{equation}
|(u*g)(p)|\leq C'_{\epsilon, A}\,\|g\|_{A,B}
e^{\beta(\epsilon|p|)}.
\label{lab27}
\end{equation}
By Lemma~3 in \cite{SmirnovSoloviev}, it follows that the functional $u$
has a unique continuous extension to ${\cal E}^\alpha_\beta$. It remains to
show that $V^*$ is a carrier cone for this extension. In the simplest case
of $V=\oR^n$ and $V^*=\{0\}$, this follows from Lemma 1. Indeed, ${\bf v}\in
{\cal A}_{\alpha_*}^{\beta^*}(\oR^n)$ is an entire function increasing no
faster than $C_\epsilon\exp\{\alpha_*(\epsilon|z|)$. Estimating the
coefficients $c_k$ of the power series that represents this function by
means of the Cauchy formula results in $|c_\kappa|\leq
C_\epsilon\epsilon^{|k|}/ a_{|k|}$. The test function
$g\in {\cal E}^\alpha_\beta(\{0\})$ is also entire, and it satisfies the
inequality
$$
|g(p+iq)|\leq\|g\|_Ae^{\alpha(2A|p+iq|)},
$$
whence follows the estimate
$$
|\partial^{\,\kappa}g(0)|\leq
\|g\|_A A'^{|\kappa|}\kappa!  \inf_{s>0}s^{-k}e^{\alpha(s)}.
$$
In view of the Stirling formula, the relation (\ref{lab12}) shows that the
functional
$$
u=\sum_{\kappa\in \oN^n} i^\kappa c_\kappa\partial^{\,\kappa} \delta
$$
is defined on ${\cal E}^\alpha_\beta(\{0\})$ and is continuous.

The further argument repeats the proof of Theorem 4 in \cite{Soloviev3}.
Let us consider the case of $V= {\oR}_+$ and $V^* = \bar{{\oR}}_+ $. We
apply Theorem 5 and decompose $u$ into the sum of the functionals $u_+$ and
$u_-$ carried by the respective half-axes $\bar{{\oR}}_+$ and
$\bar{{\oR}}_-$. Performing the Laplace transformation, we obtain
${\bf v}(x+i0) ={\bf v}_+(x+i0) + {\bf v}_-(x-i0)$, where ${\bf v}_\pm \in
{\cal A}_{\alpha_*}^{\beta^*}({\oC}_\pm)$. By the ``edge-of-the-wedge''
theorem, there is an entire function serving as an extension for both
${\bf v}-{\bf v}_+$ and ${\bf v}_-$. It increases no faster than
$C_\epsilon\exp\{\alpha_*(\epsilon|z|)$, and the origin is therefore a
carrier cone of the functional $u-u_+$, which completes the proof for
$n=1$. In the general case, it can be assumed that the first coordinate
unit vector lies in the cone $V$. We take
$g \in {\cal E}^\alpha_\beta({\oR}^{n-1})$, write $x' = (x_2,\ldots,x_n)$,
and consider the mapping
$$
g\rightarrow {\bf v}_1 (z_1) = \int {\bf v}(z_1, x')\check g(x')\,{\rm d}x'
$$
of ${\cal E}^\alpha_\beta({\oR}^{n-1})$ into ${\cal
  A}_{\alpha_*}^{\beta^*}({\oC}_+)$. It is easy to see that it is
continuous. Let $u_1\in {\cal E}^{\prime \alpha}_\beta (\bar{{\oR}}_+)=
{\cal E}^{\prime \alpha}_\beta ({\oR}_+)$ be the functional that is taken
to ${\bf v}_1$ by the Laplace transformation. The correspondence
${\bf v}_1\rightarrow u_1$ is also continuous by the open mapping theorem
(see Sec. 5.6 in \cite{ReedSimon}). We thus obtain a bilinear separately
continuous functional on ${\cal
E}^\alpha_\beta({\oR}_+)\times {\cal E}^\alpha_\beta({\oR}^{n-1})$, which
determines by Theorem 3 a continuous linear functional on ${\cal
E}^\alpha_\beta(H_1)$, where $H_1$ is the half-space $x_1>0$.
The restriction of this functional to ${\cal E}^\alpha_\beta({\oR})\otimes
{\cal E}^\alpha_\beta({\oR}^{n-1})$ coincides with that of the functional
$u$, and we conclude that $\bar H_1$ is a carrier cone for $u$. The same is
true for each of the half-spaces $\{p:  py\geq 0\}$ с $y \in V$ with
$y\in V$, whose intersection is the cone $V^*$, which consequently is also
a carrier cone of the functional $u$ by Theorem 6. This completes the
proof.

It is also useful to have a theorem characterizing spectral functions in
the case of preliminarily unknown regularity properties for the growth of
holomorphic functions near the real boundary of the analyticity domain. The
statement below is suitable for our aims.

{\bf Theorem 9.} {\it
Suppose ${\bf v}$ is a holomorphic function in $T^V$, and for any compact
subcone $V'$ of $V$ and for an arbitrary $\epsilon>0$ the inequality
\begin{equation}
|{\bf v}(z)|\leq
C_\epsilon(V')\exp\{\alpha_*(\epsilon|z|)+\gamma_{V'}(|y|)\}\quad
(y\in V')
\label{lab28}
\end{equation}
holds, where $\gamma_{V'}$ is a nonnegative
function indefinitely increasing with decreasing argument and
$\alpha_*$ is conjugate to a function $\alpha$ possessing the
properties specified in Definition $1$ and increases no faster than a
linear exponential function. Then ${\bf v}$ is the Laplace transform of an
analytic functional belonging to any space ${\cal
  E}^{\prime\alpha}_\beta(V^*)$, where $\beta$ satisfies $(\ref{lab20})$ and the
inequality
\begin{equation}
(-\gamma_{V'})^*(s)\leq C'_\epsilon(V')+\beta(\epsilon s).
\label{lab29}
\end{equation}
}
\indent Indeed, substituting (\ref{lab28})
in (\ref{lab25}), we obtain an estimate of the
type (\ref{lab26}) with $\gamma_{V'}(t)$ instead of $-\beta^*(t/\epsilon)$.
As before, the monotonicity of $\gamma_{V'}$ permits passing to the infimum
throughout the half-axis in this estimate, which yields the inequality
(\ref{lab27})
by virtue of (\ref{lab29}). The remaining part of the proof is the
same as in
Theorem 8 in view of the inequality $\gamma_{V'}(t)\leq
C'_\epsilon(V')-\beta^*(t/\epsilon)$ following from (\ref{lab29}).

\section{Generalized spectral condition}

The functional domain of definition of the Wick series
$\sum d_k:\phi^k:(x)$ was found in \cite{SmirnovSoloviev} proceeding from
the infrared and ultraviolet behavior of the Hilbert majorant of the vacuum
expectation value $w(x-x')=\langle\Psi_0,\phi(x)\phi(x')\Psi_0 \rangle$.
The majorant is the boundary value of a function ${\bf w}_{{\rm maj}}(z,
 z')$ holomorphic in the tubular domain $\{(z,z')\in \oC^{\tt 2d}:\, y=
{\rm Im\,}z\in \oV_-,\, y'= {\rm Im\,}z'\in \oV_+\}$, and its behavior can
be characterized by the inequality
\begin{equation}
|{\bf w}_{{\rm maj}}(z, z')|\leq C_0+C_1\,w_{{\scriptscriptstyle IR}}
(|z|+|z'|)+C_2\,w_{{\scriptscriptstyle UV}}(|y|+|y'|),
\label{lab30}
\end{equation}
where the pair $(y,y')$ runs through a compact subcone of
$\oV_-\times \oV_+$, the constant $C_2$ depends on this subcone, and
$w_{{\scriptscriptstyle IR}}$ and $w_{{\scriptscriptstyle UV}}$ are
nonnegative monotonic functions, the first of them increasing, and the
other decreasing. The inequality (\ref{lab30}) was used in \cite{SmirnovSoloviev}
only for $y$ and $y'$ belonging to the $y_0$-axis, whereas here we need its
satisfaction to the full extent.

Let the series $\sum d_k:\phi^k:(x)$ be convergent under averaging with
test functions belonging to $S^b_a={\cal F}({\cal E}^\alpha_\beta)$. By
the {\it generalized spectral condition}, we mean the requirement that the
closed cone (\ref{lab5})
be a carrier cone of the vector-valued analytic functional
(\ref{lab2}) in the momentum representation. Let $\check f\in
{\cal E}^\alpha_\beta(K_{n-})$, let $\check f_\nu\in{\cal
E}^\alpha_\beta$, and let $\check f_\nu\to \check f$. To prove the
validity of the stated condition, it suffices to show that
$\Psi(f_\nu)=\hat\Psi(\check f_\nu)$ is a Cauchy sequence. This is true if
the acute cone $(-K_{n-})\times K_{n-}$ is a carrier cone of the Fourier
transform of the generalized function $T_n$ in the right-hand side of
(\ref{lab1}).
Theorem 9 reduces deriving this property to estimating the behavior of the
corresponding analytic function ${\bf T}_n(z)$. It should be taken into
consideration here that the transition from the momentum representation to
the coordinate representation in quantum field theory is usually realized
with the opposite sign in the Fourier--Laplace transformation as compared
with the one used in the foregoing section. Explicitly, we have
${\bf T}_n(z)=\sum_KD_K\,{\bf W}^K(z)$, where $z\in
 \oC^{2n{\tt d}}$, $K$ is a multi-index with the components $k_{jm}$,
the coefficients $D_K$ can be expressed via $d_k$ in a known way, and
$$
{\bf W}^K(z)=\prod_{1\le j<m\le n}{\bf w}(z_m-z_j)^{k_{jm}}
\prod_{ n+1\le j<m\le 2n} {\bf w}(z_j-z_m)^{k_{jm}}
\prod_{1\le j\le n\atop n+1\le m\le 2n}{\bf w}_{{\rm maj}}(z_j,z_m)^{k_{jm}}.
$$
We write $V_{n-}={\rm int}\,K^*_{n-}\,$ and $V=V_{n-}\times (-V_{n-})$.
Then
$$
\begin{array}{l}
V=\{y\in\oR^{2n{\tt d}}:\, y_1\in\oV_-,\quad
y_{j}-y_{j-1}\in\oV_-\,\,(2\leq j\leq n),\\
\hspace{4.7cm}
y_{n+1}\in \oV_+,\quad y_j-y_{j-1}\in\oV_+\,\, (n+2\le j\leq 2n)\}.
\end{array}
$$
Let a cone $V'$ be compact in $V$, let $1\le j\le n$, and let $n+1\le
m\le 2n$. If $y$ ranges $V'$, then the pair $(y_j,y_m)$ ranges a compact
subcone of $\oV_-\times \oV_+$, and there is a $\delta>0$ such that
$|y_j|+|y_m|\geq\delta|y|$. Indeed, let ${\rm pr}\,V'$ denote the
projection of the cone $V'$, i.e., its intersection with the unit sphere
in $\oR^{2n{\tt d}}$, and let the expression $|y_j|+|y_m|$ be regarded as
a function of $y$. This function is continuous, maps the compact set
${\rm pr}\,V'$ into a compact subset in $\oV_-\times \oV_+$, and attains
its infimum on ${\rm pr}\,V'$, which is nonzero because points with zero
components do not belong to $V$. This precisely implies the desired
assertion. A similar argument shows that for the indices
$1\le j<m\le n$ and $n+1\le j< m\le 2n$, the difference $y_j-y_m$ between
the components of the vector $y\in V'$ ranges compact subcones of $\oV_+$
and $\oV_-$ respectively, and we have $|y_j-y_m|\geq\delta|y|$, possibly
with a different, but nonzero, constant $\delta$. Applying the inequality
$|{\bf w}(x-x'-2i\eta)|^2\leq |{\bf w}_{{\rm maj}}(x-i\eta,\, x+i\eta)|\,
|{\bf w}_{{\rm maj}}(x'-i\eta,\, x'+i\eta)|$
proved in \cite{SmirnovSoloviev}, which holds for all $\eta\in \oV_+$,
using (\ref{lab30}), and taking into account the monotonicity of
$w_{{\scriptscriptstyle  IR}}$ and $w_{{\scriptscriptstyle UV}}$, we
obtain
\begin{equation}
|{\bf W}^K(z)|\leq 3^{|K|}
\left(C_0^{|K|}+C_1^{|K|}w_{{\scriptscriptstyle IR}}(2|z|)^{|K|}+
C_2(V')^{|K|}w_{{\scriptscriptstyle UV}}(\delta_{V'}|y|)^{|K|}\right)\qquad
y\in V'.
\label{lab31}
\end{equation}
We subject the coefficients of the Wick series in question to the
conditions (whose meaning was elucidated in \cite{SmirnovSoloviev})
\begin{equation}
d_k\geq0,\quad d_0=1,\quad\lim_{k\to\infty}
(k!d_k^2)^{1/k}=0,\quad d_kd_l \le CH^{k+l} d_{k+l},
\label{lab32}
\end{equation}
where $C$ and $H$ are constants. The third condition in (\ref{lab32}) ensures the
absolute and uniform convergence  of the series representing
the function ${\bf T}_n(z)$ on compact sets in $T^V$ and,
as a consequence, the
holomorphy of this function in $T^V$, whereas the fourth condition permits
deriving from (\ref{lab31}) the inequality

\begin{equation}
|{\bf T}_n(z)|\leq
C\left(\sum_{k=0}^\infty L^kk!d_{2k} w_{{\scriptscriptstyle
IR}}(2|z|)^k\right) \left(\sum_{k=0}^\infty
L^kk!d_{2k}w_{{\scriptscriptstyle UV}}(\delta|y|)^k\right),
\qquad y\in V',
\label{lab33}
\end{equation}
where $L$ and $\delta$ depend $V'$.

We can now summarize our results.

{\bf Theorem 10.} {\it Let $\phi$ be a free field acting in a
pseudo-Hilbert space ${\cal H}$. Under the conditions $(\ref{lab32})$
on the
coefficients, the field $\varphi(x)=\sum d_k:\phi^k:(x)$ is well defined
as an operator-valued distribution on every space $S^b_a$ whose indicator
functions satisfy the inequalities
$$
\sum_k
L^kk!d_{2k} w_{{\scriptscriptstyle IR}}(r)^k \le C_{L,\epsilon}\,a(\epsilon
r),\quad \inf_{t>0}\,e^{st} \sum_k L^kk! d_{2k} w_{{\scriptscriptstyle
UV}}(t)^k \le C_{L,\epsilon}\, b(\epsilon s)
$$
for an arbitrarily large $L>0$ and an arbitrarily small $\epsilon>0$, and,
in addition, $\log a(r)$ grows no faster than a linear exponential
function. If $S_b^a$ is a Gelfand--Shilov--Gurevich space and
$\beta(s)=\log b(s)$ satisfies $(\ref{lab20})$, then all conditions of the
pseudo-Wightman formalism hold for $\varphi(x)$ including the generalized
spectral condition.}

The first of these assertions was proved in \cite{SmirnovSoloviev}, and
the other can be derived from (\ref{lab33}) using Theorem 9 if
$\gamma_{V'}(t)=\ln\sum_k L_{V'}^kk!d_{2k}
w_{{\scriptscriptstyle UV}}(\delta_{V'} t)^k$ is taken.

\section{Conclusion}

The functional domain of definition of the Wick series of a free field
with an indefinite metric was found comparatively simply in
\cite{SmirnovSoloviev}. In contrast, proving that the basic conditions
of general quantum field theory are satisfied for sums of such series
requires studying the properties of analytic functionals rather
thoroughly. We believe that these efforts are justified because they not
only permit properly refining the pseudo-Wightman formalism but also, in
our opinion, provide a mathematical basis for a consistent Euclidean
formulation of quantum field models with singular infrared behavior
violating the positivity condition. The above condition on the growth of
the indicator function $a(r)$ is not essential and is motivated only by
the desire to give clearer, more concise proofs. We note that the Wick
series in question can always be realized on the spaces
$S_0^b$, which correspond to the spaces ${\cal E}^\alpha_\beta$ with
$\alpha(s)=s$. These spaces allow arbitrary infrared behavior because they
consist of functions of compact support in the coordinate representation,
and the corresponding formulation of the spectral condition is
the weakest (i.e., the most general). However, the objective of the
present paper is to accurately describe both the functional domain of
definition of Wick series and their spectral properties. The developed
technique permits solving another interesting problem completely, namely,
investigating the nonlocal extension of the Borchers equivalence classes,
in which the major difficulty is related to correctly generalizing the
microcausality condition and proving that it is satisfied for nonlocal
Wick series. These results will be presented in a separate publication.

{\bf Acknowledgments.} This work was supported by the Russian Foundation
for Basic Research (Grant No. 99-02-17916) and INTAS (Grant No. 99-1-590).

\end{document}